\documentclass[reprint,aps,pre,amsmath,amssymb,floatfix,superscriptaddress,
  nofootinbib,showkeys]{revtex4-2}

\usepackage[T1]{fontenc}
\usepackage[utf8]{inputenc}
\usepackage{txfonts}
\usepackage{graphicx}
\usepackage{booktabs}
\usepackage{overpic}
\usepackage{placeins}
\usepackage[colorlinks=true,linkcolor=black,citecolor=blue,urlcolor=blue]{hyperref}

\AtBeginDocument{\renewcommand{\doi}[1]{doi:\href{https://doi.org/#1}{#1}}}

\begin{document}

\title{The preplasma origin of the plasma-mirror reflectivity response}

\author{Zhongwei Wang}
\author{Douglass Schumacher}
\email{schumacher.60@osu.edu}
\affiliation{Department of Physics, The Ohio State University, Columbus,
  Ohio 43210, USA}

\begin{abstract}
    Plasma mirrors (PM) enhance the temporal contrast of high-power lasers, but declining reflectivity at subrelativistic intensities
    limits their energy throughput. Here we show that prepulse-generated preplasma drives this decline in p polarization
    primarily through absorption. Particle-in-cell simulations using a measured prepulse profile reproduce the reflectivity
    decrease, with near- and far-field energy accounting showing a weak contribution from scattering. Optical-cycle-resolved electron tracking reveals successive electron bunches extracted from the plasma surface, supporting a vacuum-heating interpretation. Under the simulated conditions, s-polarized absorption remains weak and no comparable decline occurs.
    Applying the model to two representative double-plasma-mirror (DPM) systems, BELLA and CoReLS, it captures both the p-polarized and s-polarized reflectivity responses across the operating range. Yet, the model still misses the high-fluence s-polarized reflectivity rollover, and further investigation of the scattering channel might be needed. These findings connect prepulse history to polarization-dependent energy loss, informing the choice of plasma-mirror operating conditions.
\end{abstract}

\maketitle

\section{Introduction}

High temporal contrast is essential in petawatt-class laser systems, where
prepulses and pedestals can prematurely ionize solid-density targets and alter
the conditions encountered by the main pulse. Plasma mirrors exploit the
intensity-dependent optical response of a dielectric surface to suppress this
low-intensity background
\cite{zienerSpecularReflectivityPlasma2003,Doumy2004,%
caiTimeresolvedMeasurementsReflectivity2009,%
shawReflectanceCharacterizationTapebased2016,%
pooleExperimentSimulationNovel2016}. Below the ionization threshold, an
antireflection-treated target transmits most of the incident field; as the main
pulse rises, the surface ionizes and the resulting overdense plasma becomes
highly reflective. Under optimized conditions, a single plasma mirror can
improve the temporal contrast by more than two orders of magnitude while
retaining most of the main-pulse energy
\cite{scottOptimizationPlasmaMirror2015}. For still higher contrast, two plasma
mirrors can be operated in series as a double-plasma-mirror (DPM) system
\cite{wittmannUltrahighcontrastUltraintenseLaser2006,%
levyDoublePlasmaMirror2007,kimSpatiotemporalCharacterizationDouble2011}.
Renewable liquid-crystal films have further enabled DPM operation at the
repetition rates required by modern petawatt-class facilities
\cite{Obst_2018,czaplaRenewableDoublePlasma2025}.

The useful operating range of a plasma mirror is nevertheless limited by the
non-monotonic dependence of its reflectivity on laser intensity. Measurements
show that the specular reflectivity first rises as the target ionizes, reaches
a maximum in the subrelativistic regime, and then decreases at intensities of
approximately $10^{16}$--$10^{17}$~W/cm$^2$. The location and magnitude of this turnover depend on target, laser and irradiation conditions, such as material ionization response and laser temporal contrast. Despite this variation, the turnover remains a robust
feature across widely different plasma-mirror configurations. As summarized in
Table~\ref{tab:experiments}, it has been reported for targets ranging
from fused silica and glass to VHS tape, liquid-crystal films, and polymer
foils, over a broad range of incidence angles and focusing geometries
\cite{zienerSpecularReflectivityPlasma2003,%
caiTimeresolvedMeasurementsReflectivity2009,%
shawReflectanceCharacterizationTapebased2016,%
pooleExperimentSimulationNovel2016,SpecularReflectivityDiffuse}.
Although the drop is consistently
observed in the reported p-polarized measurements, the s-polarized
response is less universal, ranging from a clear decrease to saturation.
This difference hints that an additional absorption channel may contribute
for p polarization, while the conditional s-polarized decrease points to
other loss mechanisms, most likely diffuse scattering.

\begin{table*}[t]
\caption{Measurements of plasma-mirror specular reflectivity
versus intensity}
\label{tab:experiments}
\centering
\footnotesize
\setlength{\tabcolsep}{4pt}
\begin{tabular*}{\textwidth}{@{\extracolsep{\fill}} l l c c c l l @{}}
\toprule
Reference & Target & Duration (fs) & $\theta$ & Pol. & $f/\#$ (FWHM at focus) & Reflectivity drop \\
\midrule
Ziener \textit{et al.}\ \cite{zienerSpecularReflectivityPlasma2003} & Fused silica & 90/500 & $6/18/45^\circ$ & --- & $f/17$ ($20\times26~\mu$m) & Yes \\
Cai \textit{et al.}\ \cite{caiTimeresolvedMeasurementsReflectivity2009} & Glass & 60/500 & $32.5^\circ$ & s, p & $f/4$ ($10\times10~\mu$m) & p only; s saturates \\
Poole \textit{et al.}\ \cite{pooleExperimentSimulationNovel2016} & Liquid crystal (8CB) & 40 & $15^\circ$ & s & $f/7$ ($7~\mu$m) & Yes (s) \\
Shaw \textit{et al.}\ \cite{shawReflectanceCharacterizationTapebased2016} & VHS tape & 40 & $45^\circ$ & s, p & $f/25$ ($25.9~\mu$m) & Yes \\
Benabdelghani \textit{et al.}\ \cite{SpecularReflectivityDiffuse} & UDMA--TEGDMA polymer & 50 & $45^\circ$ & s & $f/10$ ($\sim10~\mu$m) & s, large spot only \\
\bottomrule
\end{tabular*}
\end{table*}

Preplasma formed by prepulses and rising pedestal ahead of the peak pulse
has long been associated with this high-intensity reflectivity loss. Early
studies proposed that the evolving density gradient increases laser absorption
\cite{zienerSpecularReflectivityPlasma2003}, while time-resolved measurements
showed that the reflectivity begins to evolve progressively earlier as the
peak intensity is increased, consistent with pedestal-driven ionization well
before the arrival of the pulse maximum
\cite{caiTimeresolvedMeasurementsReflectivity2009}. Other models have instead
described the reduction of specular reflectivity through scattering from a
nonuniformly expanding critical surface, often represented by an empirical
surface-roughness attenuation factor. Such descriptions have reproduced
single- and double-plasma-mirror measurements quantitatively
\cite{scottOptimizationPlasmaMirror2015,%
shawReflectanceCharacterizationTapebased2016,%
czaplaRenewableDoublePlasma2025}, but they do not determine how much of the
lost specular energy is absorbed rather than scattered, nor do they identify
the microscopic absorption mechanism.

Particle-in-cell (PIC) simulation provides a route to resolving these
mechanisms directly, but previous calculations did not include the
pedestal-driven preplasma. Cochran \textit{et al.} reproduced the low-intensity
turn-on and high-intensity reflectivity plateau but not the intervening
reflectivity drop
\cite{cochranModelingPulsecleaningPlasma2019}. In the present work, we extend
this approach by incorporating the experimentally measured prepulse so that
the preplasma is formed before the main-pulse interaction. The calculation
uses PPT field ionization to capture the low-intensity turn-on without fitted
ionization-rate parameters and relativistic Monte Carlo binary collisions in
WarpX
\cite{perelomovIONIZATIONATOMSALTERNATING,%
zhaoMultiphotonTunnelingIonization2014,%
perezImprovedModelingRelativistic2012,vayWarpXNewExascale2018a}.
By combining near-field energy diagnostics with far-field propagation, we
further separate absorption from diffuse scattering. The simulations show that
the prepulse-driven preplasma produces a pronounced loss of p-polarized
reflectivity, whereas the s-polarized response remains largely unchanged.
The missing p-polarized energy appears predominantly as absorption rather than
scattering, and the cycle-resolved electron dynamics identify enhanced vacuum
heating at the preplasma-modified critical surface as the dominant mechanism.

We then validate whether this reflectivity model remains predictive outside the
configuration used to study the preplasma physics by comparing it with two
high-performance DPM demonstrations representative of current petawatt-class
laser systems. The first is the p-polarized BELLA experiment of Czapla
\textit{et al.}, which used renewable, free-standing 8CB liquid-crystal films
at near-normal incidence \cite{czaplaRenewableDoublePlasma2025}. The second is
the s-polarized 4-PW experiment of Choi \textit{et al.}, which used
antireflection-coated glass plasma mirrors at large incidence angle
\cite{choiHighlyEfficientDouble2020}. These benchmarks span different
polarizations, target materials, pulse durations, and interaction geometries,
providing a stringent test of the modeled plasma-mirror response. The PIC
model reproduces the broad DPM throughput dependence measured in the BELLA
system and agrees with the s-polarized measurement over most of the measured
fluence range. At the highest s-polarized fluences, however, the experiment
shows a weak reflectivity decrease that is absent from the simulation,
consistent with the finding that the present preplasma dynamics do not by
themselves generate an s-polarized reflectivity drop. Together, these results
establish the role of prepulse-driven preplasma in the p-polarized
reflectivity loss while also delineating the regime in which additional
physics is required for the s-polarized response.

\section{Methodology}\label{sec:methods}
\subsection{Ionization model and low-field validation}
\label{sub:ionization}
The Ammosov--Delone--Krainov (ADK) model
\cite{ammosovADKModel1986} is widely used for field ionization in PIC
simulations because of its simple closed-form rate \cite{vayWarpXNewExascale2018a,  smithParticlecellCodeComparison2021, derouillatSmileiCollaborativeOpensource2018}. 
However, ADK is based on the quasistatic tunneling approximation and is accurate only in the
$\gamma \ll 1$ regime, where the Keldysh parameter is
\begin{equation}
    \gamma
    = \frac{\omega\sqrt{2m_e I_p}}{eE_0}
    = \sqrt{\frac{I_p}{2U_p}},
\end{equation}

with $I_p$ the ionization potential and $U_p$ the ponderomotive energy. However, the intensity range associated with plasma-mirror turn-on lies well outside the tunneling regime where ADK is
expected to be accurate. Accurately describing ionization in this range is
particularly important here because the prepulse and rising pedestal ionize
the target at these low intensities and thereby determine the preplasma seen
by the main pulse.

\begin{figure}[!t]
    \vspace*{-1.2ex}
    \centering
    \includegraphics[width=0.9\linewidth]{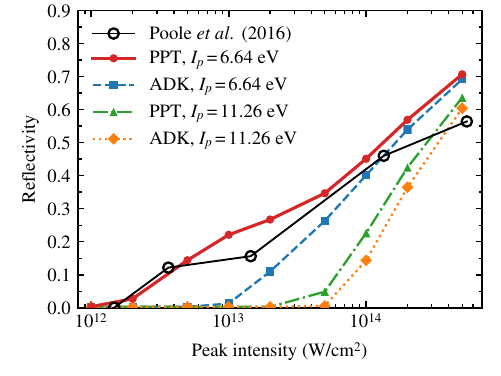}
    \caption{Low-field plasma-mirror reflectivity calculated with the PPT and
    ADK ionization models using either the DFT-calculated first ionization
    potential of 8CB, $I_p=6.64$~eV
    \cite{kumarEvolutionAnisotropyFirst2019}, or the atomic-carbon value,
    $I_p=11.26$~eV, compared with the measurement of Poole
    \textit{et al.}~\cite{pooleExperimentSimulationNovel2016}. ADK
    underestimates ionization in the high-$\gamma$ turn-on regime, while the
    larger atomic-carbon ionization potential shifts the onset to higher
    intensity. PPT with $I_p=6.64$~eV captures the measured plasma-mirror
    turn-on without fitted ionization-rate parameters.}
    \label{fig:ionization}
\end{figure}

Previous PIC modeling obtained low-field accuracy using a phenomenological
multiphoton-ionization rate with fitted parameters
\cite{cochranModelingPulsecleaningPlasma2019}. Instead, we implement the Perelomov--Popov--Terent'ev (PPT) model \cite{perelomovIONIZATIONATOMSALTERNATING,%
zhaoMultiphotonTunnelingIonization2014}, which spans the multiphoton and
tunneling regimes without fitting the ionization rate to the plasma-mirror
measurement. Direct evaluation of the PPT rate is computationally expensive
because it contains the factor $A_m(\omega,\gamma)$, expressed as an infinite
series in $\gamma$ \cite{zhaoMultiphotonTunnelingIonization2014}. We therefore
use PPT for $\gamma \geq 0.5$ and switch to the closed-form ADK expression for
$\gamma < 0.5$, where the two models converge. For the present simulations,
the laser wavelength is fixed at $\lambda=800$~nm, and the required PPT rates
are precomputed and cached so that the ionization rate can be obtained during
the PIC calculation by table lookup.

Figure~\ref{fig:ionization} validates the treatment against the measured
low-field reflectivity of the 8CB liquid-crystal plasma mirror of Poole
\textit{et al.}~\cite{pooleExperimentSimulationNovel2016}. For the same
ionization potential, ADK predicts a delayed turn-on relative to PPT,
consistent with its underestimation of ionization in the high-$\gamma$
regime. The first ionization potential is also important. To represent
ionization from the molecular HOMO of 8CB, one carbon atom per 8CB molecular
unit is assigned the DFT-calculated value $I_p=6.64$~eV
\cite{kumarEvolutionAnisotropyFirst2019}, while the remaining carbon atoms
retain the atomic-carbon value $I_p=11.26$~eV. With this molecular ionization
threshold, PPT reproduces the measured plasma-mirror turn-on without fitted
ionization-rate parameters.

\subsection{Near-field (NF) and far-field (FF) diagnostics}
\label{subsec:nf-ff}

The reduction of specular reflectivity can result from two distinct channels:
absorption or diffuse scattering. To separate these contributions, we combine near-field (NF) energy-flux diagnostics, which measure the total reflected energy, with
far-field (FF) propagation, which determines the fraction remaining within a
defined specular cone.

\subsubsection{Near-field reflectivity and absorption}

We record $\mathbf{E}(\mathbf{r},t)$ and $\mathbf{B}(\mathbf{r},t)$ on two
planes a few wavelengths in front of the target: an incident plane normal to
$\hat{\mathbf{k}}_i$ and a reflected plane normal to
$\hat{\mathbf{k}}_r$. The planes intercept the spatially separated incoming
and reflected beams. The near-field reflectivity is calculated from the
time-integrated Poynting flux,
\begin{equation}
R_{\mathrm{near}}
=
\frac{
\displaystyle \int dt\int dA\,
\mathbf{S}_r\cdot\hat{\mathbf{k}}_r
}{
\displaystyle \int dt\int dA\,
\mathbf{S}_i\cdot\hat{\mathbf{k}}_i
},
\qquad
\mathbf{S}=\frac{1}{\mu_0}\mathbf{E}\times\mathbf{B}.
\end{equation}
Because this diagnostic integrates over the full reflected beam, it includes
both specular and diffusely scattered reflected energy. Transmission
$T_{\mathrm{near}}$ is obtained analogously from a plane behind the target,
and the absorbed fraction follows from energy conservation,
\begin{equation}
A = 1-R_{\mathrm{near}}-T_{\mathrm{near}}.
\end{equation}

\subsubsection{Far-field reflectivity and scattering loss}

To isolate the specular component of the reflected energy, the reflected
near field is propagated to the far field. We first Fourier transform the
reflected electric field in time to obtain
$\widetilde{\mathbf{E}}_{r,\mathrm{near}}(x,y;\omega)$. Each frequency
component is then Fresnel propagated \cite{bornPrinciplesOpticsElectromagnetic2013}
a distance $Z$ along $\hat{\mathbf{k}}_r$,
\begin{equation}
\begin{split}
\widetilde{\mathbf{E}}_{r,\mathrm{far}}(X,Y;\omega)
={}&
\frac{e^{ikZ}}{i\lambda Z}
\int dx\,dy\,
\widetilde{\mathbf{E}}_{r,\mathrm{near}}(x,y;\omega)
\\
&\times
\exp\!\left[
\frac{ik}{2Z}
\left((x-X)^2+(y-Y)^2\right)
\right],
\end{split}
\end{equation}
where $k=\omega/c$ and $\lambda=2\pi c/\omega$. Here $(x,y)$ denote
transverse coordinates on the reflected NF plane and $(X,Y)$ those on the FF
plane.

The far-field reflectivity is defined as the reflected energy contained
within a specular aperture $\mathcal{A}$ centered on
$\hat{\mathbf{k}}_r$,
\begin{equation}
R_{\mathrm{far}}
=
\frac{
\displaystyle
\int d\omega
\int_{\mathcal{A}} dX\,dY\,
\left|
\widetilde{\mathbf{E}}_{r,\mathrm{far}}(X,Y;\omega)
\right|^2
}{
\displaystyle
\int d\omega
\int dx\,dy\,
\left|
\widetilde{\mathbf{E}}_{i}(x,y;\omega)
\right|^2
},
\end{equation}
where $\widetilde{\mathbf{E}}_i(x,y;\omega)$ is the Fourier-transformed
incident field. The normalization therefore corresponds to the same incident
energy used in $R_{\mathrm{near}}$.

There is no universal experimental definition of the angular acceptance used
to distinguish specular reflection from diffuse scattering. Reported plasma
mirror measurements, however, typically collect within a cone approximately
twice the focusing cone. Scott \textit{et al.}\
\cite{scottOptimizationPlasmaMirror2015}, for example, focused with an
$f/12.5$ cone ($\alpha\simeq2.29^\circ$) and collected the reflected light
with an $f/5.3$ optic ($5.39^\circ\simeq2.35\alpha$). Shaw \textit{et al.}\
\cite{shawReflectanceCharacterizationTapebased2016} used an $f/25$ focusing
optic ($\alpha\simeq1.15^\circ$) and an $f/13$ collection system
($2.20^\circ\simeq1.92\alpha$). We therefore use a consistent specular-cone
half-angle throughout the paper
\begin{equation}
\theta_c = 2\alpha,
\label{eq:theta_c}
\end{equation}
where $\alpha=\lambda/(\pi w_0)$ is the $1/e^2$ half-angle divergence of a
diffraction-limited Gaussian beam with waist $w_0$.

The energy reflected outside this cone is identified as diffuse scattering.
The scattering loss is therefore
\begin{equation}
S = R_{\mathrm{near}}-R_{\mathrm{far}}.
\end{equation}
The complete energy balance can consequently be written as
\begin{equation}
R_{\mathrm{far}} + S + T_{\mathrm{near}} + A = 1,
\end{equation}
allowing the decrease in specular reflectivity to be separated directly into
absorptive and scattering contributions.

\subsection{Simulation setup}
\label{subsec:setup}

The PIC calculations are performed with WarpX~26.03
\cite{vayWarpXNewExascale2018a}, extended to include the PPT
ionization model of Sec.~\ref{sub:ionization}. The electromagnetic field
is advanced with a Yee solver, and macroparticles are assigned third-order
(cubic B-spline) shape functions to suppress numerical self-heating at the
steep density gradients of the critical surface. Prior to ionization the target is treated
as a macroscopic dielectric. After ionization, binary electron--ion
Coulomb collisions are modeled with the relativistic pairwise Monte Carlo
algorithm of P\'erez \textit{et al.}
\cite{perezImprovedModelingRelativistic2012}. Neutral macroparticles are
initialized with 16 particles per cell.

The production simulations use an 800-nm Gaussian pulse with a beam waist $w_0=5~\mu$m. The 2D domain spans
$26\times18~\mu\mathrm{m}^2$ in the transverse and target-normal directions
and is discretized with $2048\times2048$ cells, corresponding to
$\Delta x\simeq12.7$~nm and $\Delta z\simeq8.8$~nm. Unless otherwise stated,
the 2D reflectivities reported below are Gaussian spot-averaged. Each 2D
production run was performed on a single NVIDIA H100 GPU and completed in
approximately one hour.

Full 3D validation uses the same pulse and $w_0=5~\mu$m geometry on a
$32\times32\times20~\mu\mathrm{m}^3$ domain with $1024^3$ cells, giving
$\Delta x=\Delta y\simeq31$~nm and $\Delta z\simeq20$~nm. The 3D simulations
were run with MPI on ten CPU nodes of the Ohio Supercomputer Center Cardinal
cluster (960 ranks) and required approximately 17~h per run.

For simulations including preplasma, we first evolve a separate 2D
prepulse calculation using the measured BELLA temporal contrast of Shaw
\textit{et al.}~\cite{shawReflectanceCharacterizationTapebased2016},
represented by the two-Gaussian fit to the picosecond shoulder and
amplified-spontaneous-emission pedestal. Only the final 3~ps before the
main-pulse peak are retained; at earlier times the contrast is below
$10^{-5}$, corresponding to intensities below $\sim10^{12}$~W/cm$^2$ for
the peak intensities considered here and producing negligible additional
ionization. The resulting species- and charge-state-resolved preplasma
distributions are revolved about the target-normal $z$ axis to construct
the 3D initial condition. This axisymmetric mapping is motivated by the approximately
axisymmetric preplasma profile obtained in the prepulse calculation and is
further supported by the close agreement between the Gaussian spot-averaged
2D and full 3D main-pulse simulations. The main pulse is then propagated through this initialized
preplasma for 400~fs.

The adequacy of this 2D reduction is established in
Fig.~\ref{fig:r-s-vs-p}(b) below, where the spot-averaged 2D scan
is compared directly with full 3D simulations.

\section{Results}\label{sec:results}

\subsection{Preplasma-driven reflectivity loss}

\begin{figure}[!t]
    \vspace*{-1.2ex}
    \centering
    \begin{overpic}[width=0.9\linewidth]{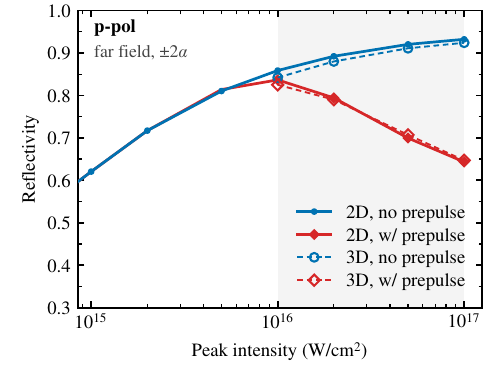}
        \put(18,16){\fontsize{9pt}{11pt}\selectfont\textbf{(a)}}
    \end{overpic}
    \par\vspace{0.3em}
    \begin{overpic}[width=0.9\linewidth]{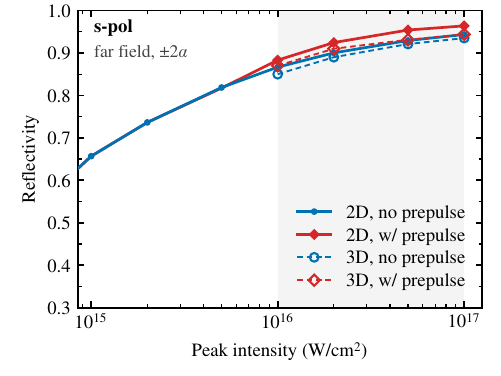}
        \put(18,16){\fontsize{9pt}{11pt}\selectfont\textbf{(b)}}
    \end{overpic}
    \caption{Influence of prepulse-driven preplasma on high-field plasma-mirror
    reflectivity. The 2D results are Gaussian-weighted over the laser spot and
    agree closely with the corresponding 3D simulations. Reflectivity is evaluated
    in the far field within $\pm 2\alpha$ of the specular direction \textbf{(a)}~p-polarization: including preplasma produces a pronounced reflectivity drop relative to the
    no-preplasma baseline. \textbf{(b)}~s-polarization: the reflectivity remains largely insensitive to preplasma over the same intensity range.}
    \label{fig:r-s-vs-p}
\end{figure}

Using the prepulse-driven preplasma described in Sec.~\ref{subsec:setup}, the high-intensity plasma-mirror response changes significantly, but the change is observed only for p polarization. Figure~\ref{fig:r-s-vs-p} shows that the preplasma produces a pronounced decrease in p-polarized reflectivity above $\sim10^{16}$~W/cm$^2$, whereas the corresponding s-polarized reflectivity remains close to the preplasma-free result. To our knowledge, this
is the first PIC reproduction of the preplasma-driven p-polarized
plasma-mirror reflectivity drop.

The Gaussian spot-averaged 2D results closely follow the corresponding 3D
simulations for both polarizations. This agreement indicates that the
polarization-selective response is not an artifact of dimensional reduction, consistent with \cite{cochranModelingPulsecleaningPlasma2019}. This motivates us to use spot-averaged 2D reflectivity instead of 3D simulations to reduce the computational cost of the rest of the work.

The reduction in specular reflectivity can in principle arise from two
distinct loss channels: absorption and scattering. We therefore
separate these contributions using the near- and far-field diagnostics
described in Sec.~\ref{subsec:nf-ff}. 

\begin{figure}[!t]
    \vspace*{-1.2ex}
    \centering
    \begin{overpic}[width=0.9\linewidth]{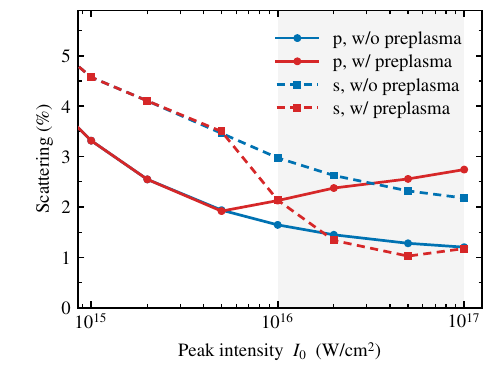}
        \put(18,67){\fontsize{9pt}{11pt}\selectfont\textbf{(a)}}
    \end{overpic}
    \par\vspace{0.3em}
    \begin{overpic}[width=0.9\linewidth]{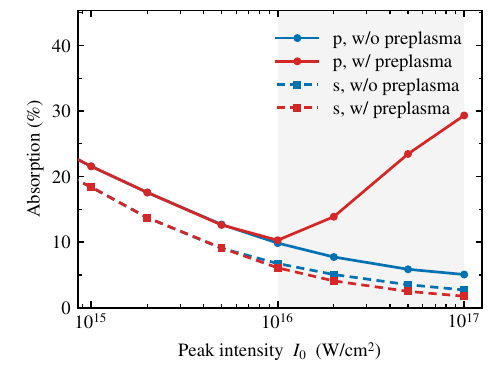}
        \put(18,67){\fontsize{9pt}{11pt}\selectfont\textbf{(b)}}
    \end{overpic}
    \caption{Decomposition of the preplasma-induced reflectivity loss into
    scattering and absorption. \textbf{(a)} Scattering remains below 5\% for both
    polarizations. The s-polarized cases exhibit
    slightly greater scattering than the corresponding p-polarized cases, while
    preplasma produces only a small, polarization-dependent change.
    \textbf{(b)} Preplasma strongly enhances p-polarized absorption at high
    intensity, from $\sim5\%$ to $\sim30\%$ at $10^{17}$~W/cm$^2$, while
    s-polarized absorption remains largely unchanged.}
    \label{fig:dpm-scatter-vs-absorption}
\end{figure}

Figure~\ref{fig:dpm-scatter-vs-absorption} shows that the two channels behave
very differently. Diffuse scattering remains below approximately 5\% over
the entire intensity range, consistent with experimental reports \cite{SpecularReflectivityDiffuse}, and does not increase with the p-polarized
reflectivity drop. In contrast, p-polarized absorption rises sharply once
the preplasma is included, reaching nearly 30\% at
$10^{17}$~W/cm$^2$. The corresponding s-polarized absorption continues to
decrease with intensity and remains only a few percent.

The missing p-polarized reflectivity therefore appears predominantly as
absorbed rather than diffusely scattered energy. Its strong polarization dependence further constrains the underlying mechanism: the enhancement requires the electric-field component normal to the plasma surface, pointing to vacuum heating. Evidence for this identification is examined in Sec.~\ref{subsec:vacuum-heating}.

\subsection{Comparison with double-plasma-mirror experiments}

We next validate the PIC model against two substantially different
double-plasma-mirror experiments performed on state-of-the-art
petawatt-class laser systems. Czapla \textit{et al.} employed
p-polarized light at near-normal incidence on renewable, free-standing
8CB liquid-crystal films at the BELLA Petawatt Laser Center
\cite{czaplaRenewableDoublePlasma2025}, whereas Choi \textit{et al.}
used s-polarized light at large incidence angle on multilayer
antireflection-coated glass plasma mirrors in a 4-PW system
\cite{choiHighlyEfficientDouble2020}. Table \ref{tab:dpm-validation} shows their experimental configurations. Agreement across these distinct platforms could provide a strong test of our model.

\begin{table}[t]
\centering
\caption{Experimental DPM configurations used for model validation.}
\label{tab:dpm-validation}
\footnotesize
\setlength{\tabcolsep}{4pt}
\begin{tabular}{@{}l c c c l@{}}
\toprule
Experiment & Pol. & $\theta_i$ & Pulse & Plasma mirror \\
\midrule
Czapla \textit{et al.} & p & $8^\circ/11^\circ$ & 35 fs & 8CB LC film, 20--40~nm \\
Choi \textit{et al.} & s & $50^\circ$ & 22 fs & AR-coated glass \\
\bottomrule
\end{tabular}
\end{table}

\subsubsection{p-polarized BELLA DPM}

Czapla \textit{et al.} demonstrated a renewable DPM at the BELLA
Petawatt Laser Center using two free-standing ultrathin 8CB
liquid-crystal films \cite{czaplaRenewableDoublePlasma2025}. The
p-polarized beam was incident on PM1 and PM2 at $8^\circ$ and
$11^\circ$, respectively, with the two mirrors separated by
$d=140$~mm. The experiment achieved a maximum DPM throughput exceeding
80\% while preserving the spatial mode and wavefront, providing a
stringent benchmark for the present model.

\begin{figure}[!t]
    \vspace*{-1.2ex}
    \centering
    \begin{overpic}[width=0.9\linewidth]{%
        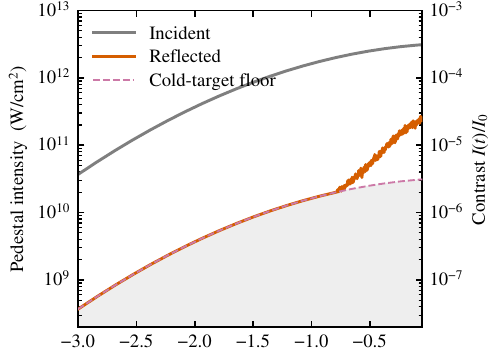}
        \put(18,73){\fontsize{9pt}{11pt}\selectfont\textbf{(a)}}
    \end{overpic}
    \par\vspace{0.3em}
    \begin{overpic}[width=0.9\linewidth]{%
        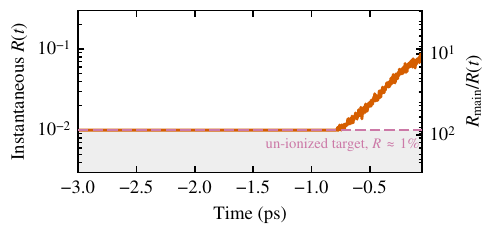}
        \put(18,42){\fontsize{9pt}{11pt}\selectfont\textbf{(b)}}
    \end{overpic}

    \caption{Time-resolved response of PM1 to the measured prepulse profile
    used in the double-plasma-mirror experiment of Czapla \textit{et al.}
    \cite{czaplaRenewableDoublePlasma2025}, for a main-pulse peak intensity
    $I_0=10^{16}$~W/cm$^2$. The cold-target floor is the dielectric
    reflectivity of the un-ionized 25-nm film, calculated using
    $n_e(\theta=\pi/12)=1.54$. \textbf{(a)} Incident and reflected pedestal
    intensity, with the corresponding contrast relative to the main-pulse
    peak shown on the right axis. The plasma mirror begins to ionize
    approximately 0.8~ps before the main-pulse peak. \textbf{(b)}
    Instantaneous plasma-mirror reflectivity $R(t)$; the right axis shows
    the corresponding ratio $R_{\mathrm{main}}/R(t)$.}
    \label{fig:pm1-pedestal-profiles}
\end{figure}

In the PIC simulations, each plasma mirror is represented by a
25-nm 8CB film with ordinary and extraordinary refractive indices
$n_o=1.53$ and $n_e=1.73$, respectively. Pulse cleaning by the first
mirror is incorporated by calculating the time-resolved reflected pulse
from PM1 and then using this modified temporal profile as the incident pulse in a second set of
simulations to obtain the PM2 reflectivity $R_2$. Figure~\ref{fig:pm1-pedestal-profiles} illustrates the cleaning at a main-pulse peak intensity of
$10^{16}$~W/cm$^2$; the cold-target floor is the dielectric
reflectivity of the un-ionized 25-nm film, evaluated with
$n_e(\theta=\pi/12)=1.54$.

To reproduce the experimental focal-position scan, the intensity on
each mirror is determined from Gaussian-beam propagation. Let $z$
denote the focal position measured downstream from PM1, with PM1 at
$z=0$ and PM2 at $z=d$. The peak intensity at the beam waist is
$I_0=8.1\times10^{17}$~W/cm$^2$, so the intensity incident on PM1 is
\begin{equation}
    I_1(z)
    =
    \frac{I_0}
    {1+\left(z/z_R\right)^2},
\end{equation}
where the Rayleigh range is
\begin{equation}
    z_R=\frac{\pi w_0^2}{\lambda}.
\end{equation}
Using $w_0=60~\mu$m and $\lambda=800$~nm gives
$z_R\simeq14.1$~mm.

\begin{figure}[!t]
    \vspace*{-1.2ex}
    \centering
    \begin{overpic}[width=0.9\linewidth]{%
        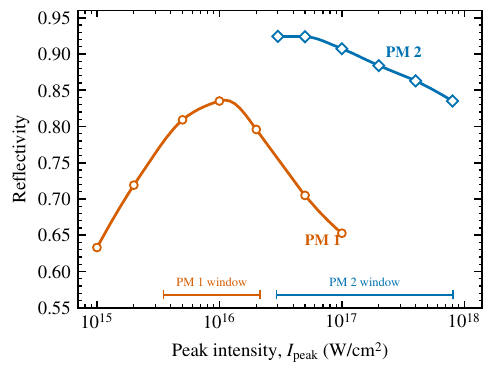}
        \put(18,78){\fontsize{9pt}{11pt}\selectfont\textbf{(a)}}
    \end{overpic}
    \par\vspace{0.3em}
    \begin{overpic}[width=0.9\linewidth]{%
        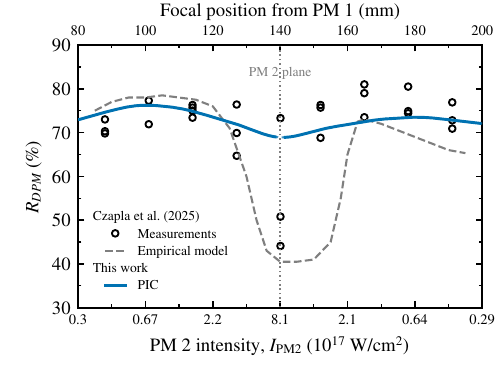}
        \put(18,73){\fontsize{9pt}{11pt}\selectfont\textbf{(b)}}
    \end{overpic}
    \caption{
    Validation against the p-polarized liquid-crystal DPM experiment of
    Czapla et al.~\cite{czaplaRenewableDoublePlasma2025}.
    \textbf{(a)} Simulated far-field reflectivity of PM1 and PM2 after pulse
    cleaning by PM1. Horizontal bars indicate the intensity ranges sampled during
    the focal-position scan. PM1 operates near maximum reflectivity, whereas PM2
    samples the decreasing high-intensity regime.
    \textbf{(b)} DPM reflectivity versus focal position relative to PM1, with the
    corresponding PM2 peak intensity shown on the lower axis. The dotted line marks
    the PM2 plane. The PIC model reproduces the broad experimental dependence,
    whereas the empirical model exhibits a narrower minimum near PM2.
    }
    \label{fig:dpm-p}
\end{figure}

After reflection from PM1, the pulse energy is reduced by
$R_1[I_1(z)]$, so the peak intensity incident on PM2 is
\begin{equation}
    I_2(z)
    =
    R_1[I_1(z)]
    \frac{I_0}
    {1+\left[(d-z)/z_R\right]^2}.
\end{equation}
The total DPM reflectivity as a function of focal position is then
\begin{equation}
    R_{\mathrm{DPM}}(z)
    =
    R_1[I_1(z)]\,R_2[I_2(z)],
    \label{eq:dpm-p}
\end{equation}
where $R_2$ is evaluated with the PM1-cleaned temporal profile rather
than the original incident pulse. The intensity-dependent
$R_1$ and $R_2$ obtained in this way are shown in
Fig.~\ref{fig:dpm-p}(a), and their combination through
Eq.~\eqref{eq:dpm-p} is mapped onto the focal-position axis in
Fig.~\ref{fig:dpm-p}(b). The PIC model reproduces the broad experimental
throughput curve, including the central dip near PM2, with a maximum
simulated reflectivity of approximately 75\% compared with the
experimental peak exceeding 80\%.

Figure~\ref{fig:dpm-p}(b) compares the PIC prediction with the
measurements of Czapla \textit{et al.}~\cite{czaplaRenewableDoublePlasma2025}
and with their empirical model, which combines a Drude--Fresnel plasma
reflectivity with a surface-expansion attenuation
\cite{shawReflectanceCharacterizationTapebased2016,scottOptimizationPlasmaMirror2015}.
Near PM2 focus, where the on-mirror intensity is highest, the measured
reflectivity exhibits substantial shot-to-shot fluctuations; the PIC curve
falls within this band, whereas the empirical model predicts a deeper
minimum.

\subsubsection{s-polarized 4-PW DPM}

As a complementary benchmark, Choi \textit{et al.} demonstrated a
high-efficiency DPM on a 4-PW Ti:sapphire laser, achieving a maximum
reflectivity of approximately 70\% together with a contrast enhancement
factor of $7\times10^5$ \cite{choiHighlyEfficientDouble2020}. In
contrast to the BELLA configuration, the experiment used s-polarized
22-fs pulses incident at $50^\circ$ on large antireflection-coated
glass plasma mirrors.  

\begin{figure}[!t]
    \vspace*{-1.2ex}
    \centering
    \begin{overpic}[width=0.9\linewidth]{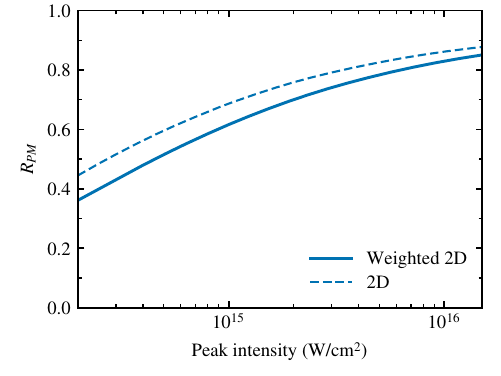}
        \put(18,67){\fontsize{9pt}{11pt}\selectfont\textbf{(a)}}
    \end{overpic}
    \par\vspace{0.3em}
    \begin{overpic}[width=0.9\linewidth]{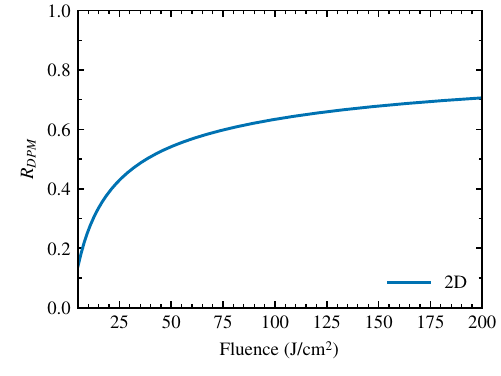}
        \put(18,67){\fontsize{9pt}{11pt}\selectfont\textbf{(b)}}
    \end{overpic}
    \caption{
    Validation against the s-polarized SiO$_2$ DPM experiment of
    Choi et al.~\cite{choiHighlyEfficientDouble2020}.
    \textbf{(a)} Simulated single-mirror far-field reflectivity for a 22-fs pulse
    at $45^\circ$ incidence, shown directly and after Gaussian spot averaging.
    \textbf{(b)} DPM reflectivity calculated as
    $R_{\mathrm{DPM}}(I)=R_1(I)R_2[I R_1(I)]$, with intensity converted to
    fluence. The spot-averaged PIC result agrees well with the measurements over
    most of the range, but does not reproduce the slight decrease above
    $\sim150$~J/cm$^2$.
    }
    \label{fig:korea-dpm}
\end{figure}

We model the Choi \textit{et al.} experiment using s-polarized light at
$45^\circ$ incidence, approximating the experimental $50^\circ$ geometry.
The plasma mirror is represented by a 300-nm SiO$_2$ layer with
$n=1.4533$, with the thickness chosen to reproduce the low-field
antireflection condition rather than explicitly resolving the multilayer
AR coating. The resulting single-mirror far-field reflectivity is shown in
Fig.~\ref{fig:korea-dpm}(a), both directly and after Gaussian spot
averaging.

Using the spot-averaged single-mirror response, the DPM reflectivity is
calculated by accounting for the energy removed by the first mirror,
\begin{equation}
    R_{\mathrm{DPM}}(I)
    =
    R_1(I)\,
    R_2\!\left[I R_1(I)\right].
\end{equation}
For a Gaussian temporal intensity profile with FWHM duration
$\tau_{\mathrm{FWHM}}$, the peak intensity is converted to fluence through
\begin{equation}
    F
    =
    I\,\tau_{\mathrm{FWHM}}
    \sqrt{\frac{\pi}{4\ln 2}}\,
    \cos\theta_i ,
    \label{eq:intensity-to-fluence}
\end{equation}
where the factor $\cos\theta_i$ accounts for the increased illuminated
area at oblique incidence. The resulting DPM reflectivity is compared with
the experimental measurements in Fig.~\ref{fig:korea-dpm}(b).

The spot-averaged PIC result agrees well with the measured reflectivity over
most of the fluence range. Above approximately $150$~J/cm$^2$, however,
the experiment shows a slight decrease in reflectivity that is absent from
the simulation. This is consistent with the s-polarized calculations above,
for which including the prepulse-driven preplasma does not produce a
high-intensity reflectivity drop.

\section{Discussion}
\label{sec:discussion}

The results above establish that the prepulse-driven preplasma produces a
strong reflectivity loss for p polarization primarily through absorption,
whereas the s-polarized response remains largely unchanged. Here we examine
the density profile formed by the measured prepulse and the cycle-resolved
electron dynamics near the critical surface to clarify the origin of this
polarization-selective absorption. We then return to the unresolved absence
of an s-polarized reflectivity drop in the simulations.

\subsection{Preplasma formed by the measured prepulse}
\label{subsec:preplasma-conditions}

To characterize the plasma encountered by the main pulse, we define the
local density scale length at the critical surface as
\begin{equation}
    L_c =
    \left|
    \frac{\partial \ln n_e}{\partial z}
    \right|_{z=z_c}^{-1},
    \qquad
    n_e(z_c)=n_c .
\end{equation}

\begin{figure}[!t]
    \centering

    \begin{overpic}[width=\linewidth]{%
        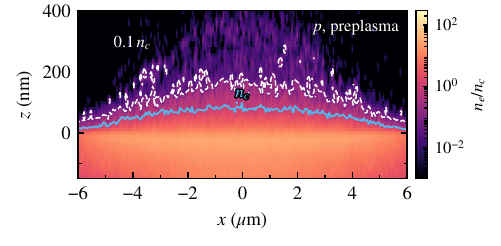}
        \put(18,43){\fontsize{9pt}{11pt}\selectfont\color{white}\textbf{(a)}}
    \end{overpic}

    \par\vspace{0.3em}

    \begin{overpic}[width=\linewidth]{%
        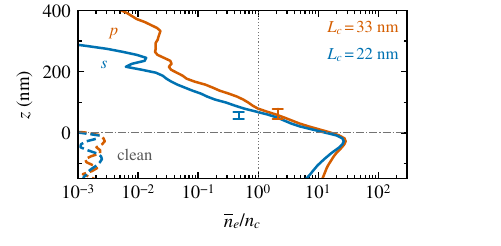}
        \put(73,17){\fontsize{9pt}{11pt}\selectfont\textbf{(b)}}
    \end{overpic}

    \caption{
    Prepulse-driven preplasma for a main-pulse peak intensity of
    $10^{17}$~W/cm$^2$ at $t=-2\tau_{\mathrm{FWHM}}$.
    \textbf{(a)} Electron-density distribution for p polarization; the
    $0.1n_c$ and $n_c$ contours mark the underdense extent and critical
    surface, respectively.
    \textbf{(b)} Electron-density profiles near the beam center for
    p- and s-polarized prepulse-driven cases and the corresponding
    clean-pulse controls. The critical-surface scale lengths are
    $L_c=33$~nm and $22$~nm for p and s polarization
    ($L_c/\lambda_0\simeq0.041$ and $0.028$), respectively.
    At the main-pulse peak they become $38.8$ and $21.5$~nm, while the
    clean-pulse scale length remains below approximately $5$~nm.
    }
    \label{fig:preplasma-conditions}
\end{figure}

Figure~\ref{fig:preplasma-conditions} shows the density profile generated by
the measured prepulse for a main-pulse peak intensity of
$10^{17}$~W/cm$^2$. At $t=-2\tau_{\mathrm{FWHM}}$, before the main pulse
reaches its peak, the target has already developed a finite underdense ramp.
Near the beam center, the critical-surface scale length is approximately
$L_c=33$~nm for p polarization and $22$~nm for s polarization. In contrast,
the clean-pulse control does not have enough ionization to form a critical surface yet.

The difference persists as the main pulse rises. At the pulse peak,
$L_c$ is approximately $38.8$~nm for p polarization and $21.5$~nm for
s polarization, whereas the clean case remains below approximately
$5$~nm until the main pulse itself ionizes the surface. The prepulse
therefore establishes the density gradient before the high-energy portion
of the pulse arrives, rather than allowing the gradient to develop only
after the interaction has already begun. For $\lambda_0=800$~nm, the
prepulse-driven profiles correspond to $L_c/\lambda_0\simeq0.041$ and $0.028$
for p and s polarization at $t=-2\tau_{\mathrm{FWHM}}$, and approximately
$0.049$ and $0.027$ at the pulse peak, respectively. Thus, despite the
prepulse-induced expansion, the interaction remains in a steep-gradient
regime. The somewhat longer p-polarized ramp provides a broader
near-critical region over which the normal electric field can extract
electrons from the plasma surface, as examined below.

\subsection{Origin of the enhanced p-polarized absorption}
\label{subsec:vacuum-heating}

\begin{figure}[!t]
    \centering

    \begin{overpic}[width=\linewidth]{%
        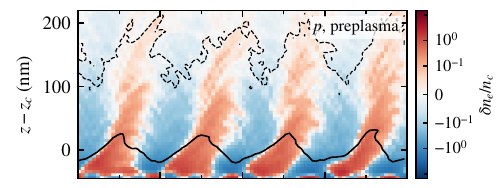}
        \put(18,31){\fontsize{9pt}{11pt}\selectfont\textbf{(a)}}
    \end{overpic}

    \par\vspace{0.3em}

    \begin{overpic}[width=\linewidth]{%
        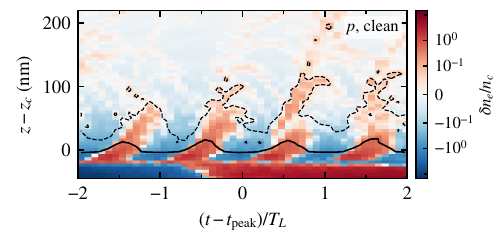}
        \put(18,43){\fontsize{9pt}{11pt}\selectfont\textbf{(b)}}
    \end{overpic}

    \caption{
    Cycle-resolved electron-density fluctuations near the critical surface
    for p-polarized irradiation around the main-pulse peak at $I_0=10^{17} \mathrm{W/cm^2}$
    \textbf{(a)} Prepulse-driven preplasma.
    \textbf{(b)} Clean-pulse control.
    In the preplasma case, electron bunches are launched from the vicinity
    of the critical surface approximately once per optical cycle and extend
    roughly $150$--$200$~nm into the underdense region. The corresponding
    excursions in the clean case remain substantially closer to the
    surface. The enhanced surface-crossing motion provides microscopic
    evidence for stronger vacuum-heating absorption in the preplasma case.
    }
    \label{fig:preplasma-density-bunches}
\end{figure}

The cycle-resolved density fluctuations in
Fig.~\ref{fig:preplasma-density-bunches} provide a microscopic view of the
enhanced p-polarized absorption. For p polarization, the electric field has
a component normal to the target surface and can therefore pull electrons
out of the overdense plasma during one part of the optical cycle and drive
them back toward the target after the field reverses. This is the
characteristic electron motion associated with vacuum, or Brunel, heating
\cite{brunelNotsoresonantResonantAbsorption1987a,
gibbonCollisionlessAbsorptionSharp1992}.

With the prepulse-driven density ramp, coherent electron-density bunches are
launched from the vicinity of the critical surface approximately once per
laser cycle and propagate into the underdense region. Their excursions
extend to roughly $150$--$200$~nm from the critical surface, and the bunches
progressively lose spatial coherence as they propagate outward. In the
clean-pulse control, the corresponding modulation is concentrated much
closer to the surface, with excursions generally below approximately
$100$~nm.

The density profiles in Fig.~\ref{fig:preplasma-conditions} provide a
natural explanation for this difference. The prepulse creates a finite but
still steep density ramp before the main pulse arrives, increasing the
spatial region over which electrons can be extracted and interact with the
normal laser field. When the field reverses, part of this electron
population is re-injected into the overdense plasma, transferring energy to
the target. The larger excursions in the preplasma case therefore provide a
cycle-resolved signature of the enhanced surface-crossing energy transfer
responsible for the increased p-polarized absorption.

Although Brunel heating is often discussed at larger incidence angles
\cite{gibbonCollisionlessAbsorptionSharp1992,
geResonantAbsorptionNotsoresonant2013,
caiShortpulseLaserAbsorption2006}, the present results show that it remains
appreciable at $15^\circ$. The normal field component is reduced but remains
finite, while $L_c/\lambda_0\sim0.04$--$0.05$ for the p-polarized preplasma
during the main interaction, so the density gradient remains sufficiently
steep for efficient surface-crossing electron motion.

\subsection{Absence of an $s$-polarized drop}
\label{subsec:s-no-drop}

Having identified preplasma-enhanced vacuum heating as the origin of the
$p$-polarized reflectivity drop, we now consider why an analogous drop is
absent for $s$ polarization. This result is not inconsistent with some experiments. Cai \textit{et al.} observed saturation rather than a
drop for $s$ polarization, while Benabdelghani \textit{et al.} found that
the drop depends on spot size. In contrast, Poole \textit{et al.} and Shaw
\textit{et al.} reported clear $s$-polarized drops
\cite{caiTimeresolvedMeasurementsReflectivity2009,SpecularReflectivityDiffuse,%
pooleExperimentSimulationNovel2016,shawReflectanceCharacterizationTapebased2016}.

The mechanism responsible for the $p$-polarized drop is not expected to
produce the same response for $s$ polarization because vacuum heating
requires an electric-field component normal to the target surface. Collisional absorption is also unlikely to
produce a substantial drop, despite the prepulse-driven increase in scale
length. For an exponential density ramp, the inverse-bremsstrahlung
absorption fraction is \cite{kruer}
\begin{equation}
    f_{\mathrm{IB}} =
    1-\exp\left[-\frac{8}{3}\frac{\nu_{ei}^{*}L_c}{c}
    \cos^3(\theta_i)\right].
\end{equation}
Using $L_c=21.5$~nm, a median electron bulk temperature of $124$~eV,
$\theta_i=15^\circ$, and $Z_{\mathrm{eff}}=4$ at
$10^{17}$~W/cm$^2$ gives $f_{\mathrm{IB}}=2.6\%$, comparable to the
absorption observed in the corresponding simulations.

A possible missing contribution is stronger diffuse scattering. The
rough-surface empirical model used in
\cite{scottOptimizationPlasmaMirror2015,
shawReflectanceCharacterizationTapebased2016,
czaplaRenewableDoublePlasma2025} requires a critical-surface roughness of
approximately $40$~nm to reproduce a $30\%$ drop in reflectivity, whereas
the present simulations produce less than $10$~nm roughness. The
tested $f/\#$ range likewise does not produce a significant change in
reflectivity. Additional transverse structure, such as realistic spatial
and temporal laser incoherence or target inhomogeneity, may therefore be
needed to reproduce the $s$-polarized drop. Such nonidealities could seed
critical-surface fluctuations that grow to amplitudes beyond those produced
by an ideal Gaussian beam incident on a smooth surface.

\section{Conclusions}

We have demonstrated how preplasma modifies the plasma-mirror reflectivity
response for both p and s polarizations. Rather than prescribing an assumed
preplasma profile, we use the measured laser temporal contrast to allow the
preplasma to form self-consistently before the arrival of the main pulse.
Accurate low-intensity ionization is essential to this approach, for which we
adopt the PPT ionization model together with the DFT-calculated first
ionization potential. Near- and far-field diagnostics are then used to
separate absorption from diffuse scattering. The resulting analysis shows
that the preplasma-induced p-polarized reflectivity loss is dominated by
absorption, while cycle-resolved electron dynamics around the critical
surface identify vacuum heating as the underlying mechanism.

The resulting reflectivity model is further validated against two representative 
high-performance DPM experiments with different target
materials, interaction geometries, pulse parameters, and polarizations. The
simulations reproduce the principal experimental trends in both configurations.
The remaining exception is the high-fluence decrease observed experimentally
for s polarization, which is not produced by the present preplasma dynamics.
This discrepancy suggests that an additional mechanism, potentially involving beam incoherence and target structure inhomogeneity, might be needed. Incorporating such nonideal
elements therefore could be the next step toward more accurate plasma-mirror modeling for petawatt-class
laser systems.

\begin{acknowledgments}
This work was supported by the U.S. Department of Energy National
Nuclear Security Administration under Contract No.~DE-NA0004202. The simulations
were performed using computational resources provided by the Ohio Supercomputer
Center \cite{ohioSupercomputerCenter1987}.
\end{acknowledgments}

\bibliographystyle{unsrtnat}
\bibliography{reflectivity-drop-p-polarization}

\end{document}